\documentclass[codesnippet]{jss}
\usepackage[latin1]{inputenc}
\usepackage{amsmath}
\usepackage{amsfonts}
\usepackage{amssymb}
\usepackage{graphicx}
\usepackage{color}
\usepackage{amsthm}

\newcommand{\xx}{\mathbf{x}}

\newcommand{\llog}{\mathrm{log}}

\newcommand{\g}{\mathbf{g}}

\newcommand{\R}{\mathbb{R}}
\newcommand{\HH}{\mathbf{H}}

\newcommand{\zz}{\mathbf{z}}
\newcommand{\SSigma}{\boldsymbol\Sigma}
\newcommand{\mmu}{\boldsymbol\mu}
\newcommand{\N}{\mathcal{N}}

\author{Alireza S. Mahani\\ Scientific Computing \\ Sentrana Inc. \And
Asad Hasan\\ Scientific Computing \\ Sentrana Inc. \AND
Marshall Jiang\\ Department of Mathematics \\ Cornell University \And
Mansour T.A. Sharabiani\\ National Heart and Lung Institute \\ Imperial College London}

\title{Stochastic Newton Sampler: \proglang{R} Package \pkg{sns}}

\Plainauthor{Alireza S. Mahani, Mansour T.A. Sharabiani} 
\Plaintitle{Stochastic Newton Sampler: R Package sns} 
\Shorttitle{Stochastic Newton Sampler: \proglang{R} Package \pkg{sns}} 

\Abstract{ 
The \proglang{R} package \pkg{sns} implements Stochastic Newton Sampler (SNS), a Metropolis-Hastings Monte Carlo Markov Chain algorithm where the proposal density function is a multivariate Gaussian based on a local, second-order Taylor series expansion of log-density. The mean of the proposal function is the full Newton step in Newton-Raphson optimization algorithm. Taking advantage of the local, multivariate geometry captured in log-density Hessian allows SNS to be more efficient than univariate samplers, approaching independent sampling as the density function increasingly resembles a multivariate Gaussian. SNS requires the log-density Hessian to be negative-definite everywhere in order to construct a valid proposal function. This property holds, or can be easily checked, for many GLM-like models. When initial point is far from density peak, running SNS in non-stochastic mode by taking the Newton step, augmented with with line search, allows the MCMC chain to converge to high-density areas faster. For high-dimensional problems, partitioning of state space into lower-dimensional subsets, and applying SNS to the subsets within a Gibbs sampling framework can significantly improve the mixing of SNS chains. In addition to the above strategies for improving convergence and mixing, \pkg{sns} offers diagnostics and visualization capabilities, as well as a function for sample-based calculation of Bayesian predictive posterior distributions.
}
\Plainkeywords{monte carlo markov chain, metropolis-hastings, newton-raphson optimization, negative-definite hessian, log-concavity} 

\Address{
Alireza S. Mahani\\
Scientific Computing Group\\
Sentrana Inc.\\
1725 I St NW\\
Washington, DC 20006\\
E-mail: \email{alireza.mahani@sentrana.com}\\
}

\usepackage{Sweave}
\begin{document}
\Sconcordance{concordance:SNS.tex:SNS.Rnw:%
1 102 1 1 0 4 1 1 4 103 1 1 2 1 0 1 1 3 0 1 2 2 1 1 8 10 0 2 2 %
1 0 5 1 1 2 4 0 2 2 29 0 1 2 7 1 1 2 1 0 1 4 6 0 2 2 1 0 4 1 3 %
0 2 2 1 0 1 1 3 0 1 2 1 3 2 0 2 1 11 0 1 2 1 3 5 0 2 2 4 0 1 2 %
1 1 1 -3 1 7 4 1 1 2 35 0 1 2 2 1 1 3 2 0 1 1 3 0 1 2 1 3 5 0 %
1 2 2 1 1 3 2 0 1 2 4 0 2 2 1 0 1 1 3 0 1 4 12 0 1 1 12 0 1 2 %
4 1 1 2 1 0 4 1 1 2 1 0 1 1 3 0 1 4 28 0 1 2 1 1 1 4 3 0 1 1 3 %
0 1 3 35 0 1 2 1 1 1 2 1 0 2 1 3 0 1 2 27 1}



\section{Introduction}\label{section-introduction}
In most real-world applications of Monte Carlo Markov Chain (MCMC) sampling, the probability density function (PDF) being sampled is multidimensional. Univariate samplers such as Slice Sampler~\citep{neal2003slice} and Adaptive Rejection Sampler~\citep{gilks1992adaptive} can be embedded in the Gibbs sampling framework~\citep{geman1984stochastic} to sample from multivariate PDFs~\citep{mahani2014mfusampler}. Univariate samplers generally have few tuning parameters, making them ideal candidates for black-box MCMC software such as JAGS~\citep{plummer-jags} and OpenBUGS~\citep{thomas2006making}. However, they become less effective as PDF dimensionality rises and dimensions become more correlated~\citep{girolami2011riemann}. Therefore, development - and software implementation - of efficient, black-box multivariate MCMC algorithms is of great importance to widespread application of probabilistic models in statistics and machine learning. Recent theoretical~\citep{girolami2011riemann,hoffman2014no} and software development~\citep{stan-software:2014} efforts to make multivariate samplers such as Hamiltonian Monte Carlo (HMC)~\citep{duane1987hybrid} more self-tuning and efficient can be viewed in this light.

The \proglang{R} package \pkg{sns} implements Stochastic Newton Sampler (SNS), a Metropolis-Hastings MCMC algorithm~\citep{hastings1970monte}, where the proposal distribution is a locally-fitted multivariate Gaussian resulting from second-order Taylor series expansion of the log-density. In its current implementation, SNS requires the log-density to be twice-diffenrentiable and globally concave, or equivalently that its Hessian matrix be negative-definite everywhere. For many Generalized Linear Models (GLMs) these conditions are satisfied~\citep{gilks1992adaptive}, and the invariance theorem of \cite{mahani2015expander} allows Hessian negative-definiteness to be studied and proven in the much lower-dimensional space of base distributions, rather than the high-dimensional space of regression coefficients.

SNS has appeared in the literature under several variations and labels. \cite{gamerman1997sampling} extend Iterative Reweighted Least Squares (IRLS) - the primary estimation technique for GLM models - to MCMC sampling by adding a Metropolis-Hastings step to it. Given that IRLS is a close cousin of Newton-Raphson optimization, their method can be considered a specialization of SNS for GLM models. \cite{qi2002hessian} present what they call `Hessian-based Metropolis-Hastings' (HMH), which is nearly identical to SNS, but they do not address the high-dimensional mixing problem, nor do they provide an open-source software implementation. More recently, the simplified manifold Metropolis adjusted Langevin Algorithm (MMALA) of \cite{girolami2011riemann} is very similar to SNS with the addition a tunable step size, or learning rate. The software accompanying their paper is written in \proglang{MATLAB}~\citep{MATLAB:2014}. The \proglang{R} package \pkg{sns}, to our knowledge, is the first open-source implemnentation of the SNS algorithm, including extensions for improving convergence (\code{rnd} and \code{nnr} arguments) and mixing (\code{part} argument), diagnostic and visualization methods (\code{summary.sns} and \code{plot.sns}), and sample-based prediction (\code{predict.sns}).

The paper is organized as follows. In Section~\ref{section-theory}, we review the theoretical background for SNS, including an overview of Metropolis-Hastings algorithms, followed by the multivariate Gaussian proposal PDF used in SNS. In Section~\ref{section-implementation}, we discuss the implementation of SNS algorithm in the \pkg{sns} package. Section~\ref{section-using} offers several examples to illustrate the usage of \pkg{sns}. Finally, Section~\ref{section-summary} provides a summary and a discussion of future research and development directions.

\section{Theory}\label{section-theory}
We begin with a brief overview of the Metropolis-Hastings algorithm.

\subsection{Metropolis-Hastings algorithm}\label{subsection-mh}
In Metropolis-Hastings (MH) MCMC sampling of the PDF, $p(\zz)$, we generate a sample $\zz^*$ from the proposal density function $q(\zz | \zz^\tau)$, where $\zz^\tau$ is the current state. We then accept the proposed state $\zz^*$ with probability $A(\zz^*, \zz^\tau)$, where:
\begin{equation}
A(\zz^*, \zz^\tau) = \mathrm{min} \left( 1 \,,\, \frac{p(\zz^*) q(\zz^\tau | \zz^*)}{p(\zz^\tau) q(\zz^* | \zz^\tau)} \right)
\end{equation}
The MH transitions satisfy detailed balance:
\begin{equation}
\begin{array}{lcl}
p(\zz)q(\zz|\zz')A(\zz',\zz) &=& \mathrm{min}(p(\zz)q(\zz|\zz'),p(\zz')q(\zz'|\zz)) \\
&=& \mathrm{min}(p(\zz')q(\zz'|\zz),p(\zz)q(\zz|\zz')) \\
&=& p(\zz')q(\zz'|\zz)A(\zz,\zz')
\end{array}
\end{equation}
The detailed balance property ensures that $p(\zz)$ is invariant under MH transitions. For a discussion of ergodicity of MH algorithm, see~\cite{roberts1999convergence}.

\subsection{SNS proposal density}\label{subsection-proposal}
SNS proposal density is a multivariate Gaussian fitted locally to the density being sampled, using the second-order Taylor-series expansion of the log-density:
\begin{equation}\label{equation-taylor}
f(\xx) \approx f(\xx_0) + \g(\xx_0)^\top \, (\xx-\xx_0) + \frac{1}{2} (\xx-\xx_0)^\top \, \HH(\xx_0) \, (\xx-\xx_0)
\end{equation}
where $f:\R^K \rightarrow \R$, and $\g$ and $\HH$ are the gradient vector and Hessian matrix for $f$, respectively, of dimensions $K$ and $K \times K$. Assuming that $f$ is globally concave, the above approximation is equivalent to fitting the following multivariate Gaussian (which we refer to as $F(\xx)$) to the PDF:
\begin{equation}\label{equation-gauss}
F(\xx) = \frac{1}{(2\pi)^{K/2}|\SSigma|^{1/2}} \exp\left\{ -\frac{1}{2}(\xx-\mmu)^T \SSigma^{-1}(\xx-\mmu) \right\}
\end{equation}
By comparing Equations~\ref{equation-taylor} and \ref{equation-gauss}, we see that the precision matrix is the same as negative Hessian: $\SSigma^{-1}=-\HH(\xx_0)$. The mean of the fitted Gaussian maximizes its log, and therefore:
\begin{equation} \label{equation-newton-step}
\mmu = \xx_0 - \HH^{-1}(\xx_0) \, \g(\xx_0)
\end{equation}
We can now formally define the (multivariate Gaussian) proposal density $q(. \,|\, \xx)$ as:
\begin{equation} \label{equation-proposal}
q(. \,|\, \xx) = \N(\xx - \HH^{-1}(\xx) \, \g(\xx) \,,\, -\HH^{-1}(\xx))
\end{equation}
Note that Equation~\ref{equation-newton-step} is simply the full Newton step~\citep{nocedal2006book}. We can therefore think of SNS as the stochastic counterpart of Newton-Raphson (NR) optimization. In NR optimization, we select the mean of the fitted Gaussian as the next step, while in SNS we draw a sample from the fitted Gaussian and apply MH test to accept or reject it. Also, note that in the special case where the sampled PDF is Gaussian, $f(\xx)$ is quadratic and therefore the proposal function is identical to the sampled PDF. In this case $A(\zz',\zz)$ is always equal to $1$, implying an acceptance rate of $100\%$.

\section{Software implementation and features}\label{section-implementation}
This section describes the components of \pkg{sns}, beginning with an overview of the SNS algorithm.
\subsection{Overview}\label{subsection-sns-algorithm}
The workhorse of \pkg{sns} package is the \code{sns} function, responsible for implementation of MH algorithm using the multivariate Gaussian proposal density described in Section~\ref{subsection-proposal}. \code{sns} implements the following steps:
\begin{enumerate}
\item
Evaluate the log-density function and its gradient and Hessian at $\xx_\text{old}$: $f_\text{old},\g_\text{old},\HH_\text{old}$.
\item\label{step-fit-1}
Construct the multivariate Gaussian proposal function at $q(.|\xx_{old})$ using Equation~\ref{equation-proposal} and $\xx=\xx_{old}$.
\item
Draw a sample $\xx_{prop}$ from $q(.|\xx_{old})$, and evaluate $logq_{prop}=\llog(q(\xx_{prop}|\xx_{old}))$.
\item
Evaluate the log-density function and its gradient and Hessian at $\xx_{prop}$: $f_{prop},\g_{prop},\HH_{prop}$.
\item\label{step-fit-2}
Construct the multivariate Gaussian proposal function at $q(.|\xx_{prop})$ using Equation~\ref{equation-proposal} and $\xx=\xx_{prop}$, and evaluate $logq_{old}=\llog(q(\xx_{old}|\xx_{prop}))$.
\item
Calculate the ratio $r=\exp((f_{prop}-f_{old})+(logq_{old}-logq_{prop}))$.
\item
If $r \geq 1$ accept $\xx_{prop}$: $\xx_{new} \leftarrow \xx_{prop}$. Else, draw a random deviate $s$ from a uniform distribution over $[0,1)$. If $s<r$, then accept $\xx_{prop}$: $\xx_{new} \leftarrow \xx_{prop}$, else reject $\xx_{prop}$: $\xx_{new} \leftarrow \xx_{old}$.
\end{enumerate}

Fitting the multivariate Gaussian in steps~\ref{step-fit-1} and ~\ref{step-fit-2} is done via calls to the private function \code{fitGaussian}. We use the functions \code{dmvnorm} and \code{rmvnorm} from package \pkg{mvtnorm} to calculate the log-density of, and draw samples from, multivariate Gaussian proposal functions.

There are two important arguments in \code{sns}, namely \code{rnd} and \code{part}. The first argument, \code{rnd}, controls whether the algorithm should run in stochastic or MCMC mode (which is the default choice), or in non-stochastic or Newton-Raphson (NR) mode. The second argument, \code{part}, controls the state space partitioning strategy. These arguments and their roles are described in Section~\ref{subsection-convergence-mixing}.

\code{sns.run} is a wrapper around \code{sns}, and offers the following functionalities:
\begin{enumerate}
\item Convenience of generating multiple samples via repeated calls to \code{sns}. After the first call, the Gaussian fit object (attached as attribute \code{gfit} in the returned value from \code{sns}) is fed back to \code{sns} via argument \code{gfit}, in order to avoid unnecessary fitting of the proposal function at current value.
\item Collecting diagnostic information such as log-probability (time series), acceptance rate, relative deviation from quadratic approximation (time series), and components of MH test. These diagnostic measures are discussed in Section~\ref{subsection-diagnostics}, and their use is illustrated via examples in Section~\ref{section-using}.
\end{enumerate}

The generic methods \code{summary.sns}, \code{plot.sns} and \code{predict.sns} provide diagnostic, visualization, and prediction capabilities, discussed in Sections~\ref{subsection-diagnostics} and \ref{subsection-prediction}, and illustrated via examples in Section~\ref{section-using}.

\subsection{Improving convergence and mixing}\label{subsection-convergence-mixing}
\textbf{NR mode:} Far from the distribution mode, the local multivariate Gaussian fit can be severely different from the PDF, leading to small overlap between the two, low acceptance rate and hence bad convergence. This can be overcome by spending the first few iterations in non-stochastic or NR mode, where instead of drawing from the proposal function we simply accept its mean as the next step. Rather than taking a full Newton step, we have implemented line search~\citep{nocedal2006optim} to ensure convergence to the PDF maximum. To use \code{sns} in NR mode, users can set the argument \code{rnd} to \code{FALSE}. In NR mode, each iteration is guaranteed to increase the log-density. Using the NR mode during the initial burn-in phase is illustrated in Section~\ref{section-using}. In \code{sns.run}, the argument \code{nnr} controls how many initial iterations to be performend in NR mode.

\textbf{State space partitioning:} Even when near the PDF maximum, the fitted Gaussian can be severely different from the PDF. This can happen if the PDF has a significant third derivative, a phenomenon that we have observed for high-dimensional problems, especially when the number of observations is small. To improve bad mixing in high dimensions, we use a strategy which we refer to as `state space partitioning', where state space is partitioned into disjoint subsets and SNS is applied within each subset, wrapped in Gibbs sampling. This functionality is available via the \code{part} argument, which is a list containing the set of state space dimensions belonging to each subset. Convenience functions \code{sns.make.part} and \code{sns.check.part} allow users to easily create partition lists and check their validity, respectively.

\subsection{Diagnostics}\label{subsection-diagnostics}
\pkg{sns} includes a rich set of diagnostics which can be accessed via functions \code{summary.sns} and \code{plot.sns}. Some of these are generic measures applicable to all MCMC chains, some are specific to MH-based MCMC algorithms, and some are even further specialized for SNS as a particular flavor of MH. Where possible, we have used the library \pkg{coda} and adopted its naming conventions, but opted to create and maintain an independent set of functions due to their specialized and extended nature.

\textbf{MCMC diagnostics:} In \code{summary.sns}, we calculate the usual MCMC chain summaries including mean, standard deviation, quantiles, and effective sample size. We also calculate a sample-based p-value for each coordinate. In \code{plot.sns} we have log-density trace plot, state vector trace plots, effective sample size by coordinate, state vector histograms, and state vector autocorrelation plots.

\textbf{MH diagnostics:} In \code{summary.sns}, we calculate the acceptance rate of MH transition proposals. If \code{mh.diag} flag is set to \code{TRUE}, all 4 components of the MH test (\code{log.p}, \code{log.p.prop}, \code{log.q} and \code{log.q.prop}) are returned as well.

\textbf{SNS diagnostics:} In \code{summary.sns}, we return \code{reldev.mean} (if \code{sns.run} was called with \code{mh.diag} set to \code{TRUE}), defined as the average relative deviation of log-density change (with respect to PDF maximum) from quadratic approximation (also constructed at PDF maximum). The location of PDF maximum is extracted from the Gaussian fit in the last iteration under NR mode. The higher this value, the more likely it is for the SNS to exhibit bad mixing. This is illustrated in Section~\ref{section-using}. For \code{reldev.mean} to be valid, the user must ensure that the value of the argument \code{nnr} supplied to \code{sns.run} is sufficiently high to ensure convergence by the end of NR phase.

\subsection{Full Bayesian Prediction}\label{subsection-prediction}
The function \code{predict.sns} allows for full Bayesian prediction, using a sample-based representation of predictive posterior distribution. It accepts an arbitrary function of state vector as argument \code{fpred}, and applies the function across all samples of state vector, supplied in the first argument, which must be an output of \code{sns.run}. The core philosophy in full Bayesian prediction is to postpone summarization of samples until the last step. For example, rather than supplying the expected values of coefficients into a function, we supply the samples and take the expected value after applying the function. Following this proper approach is important because:
\begin{enumerate}
\item Mathematically, an arbitrary function is not commutable with the expected value operator. Therefore, applying expected value early produces incorrect results.
\item For a similar reason, confidence intervals cannot be propagated through arbitrary functions. Therefore, correct uncertainty measurement also requires a full Bayesian approach.
\end{enumerate}

\section[]{Using \pkg{sns}}\label{section-using}
In this section, we illustrate how \pkg{sns} can be used via several examples. First, we launch an R session and load the \pkg{sns} package as well \pkg{mvtnorm} (used for evaluating the multivariate Gaussia log-density in example 1): 
\begin{Schunk}
\begin{Sinput}
R> library(sns)
R> library(mvtnorm)
\end{Sinput}
\end{Schunk}

\subsection{Example 1: Multivariate Gaussian}\label{subsection-example-gaussian}
Using \pkg{sns} to sample from a multivariate Gaussian is a contrived, but pedagogical, example. Since log-density for a multivariate Gaussian is quadratic, its second-order Taylor series expansion is not approximate but exact. In other words, the proposal function becomes location-independent, and equal to the sampled distribution. This means that 1) the MH test is always accepted, and 2) consecutive samples are completely independent, and hence the resulting chain is no longer Markovian. Of course, since we know how to sample from multivariate Gaussian proposal functions, we might as well directly sample from the multivariate Gaussian distribution. (Hence, the pedagogical nature of this example.) To utilize \pkg{sns}, we must first implement the log-density and its gradient and Hessian:
\begin{Schunk}
\begin{Sinput}
R> logdensity.mvg <- function(x, mu, isigsq) {
+    f <- dmvnorm(x = as.numeric(x)
+      , mean = mu, sigma = solve(isigsq), log = TRUE)
+    g <- - isigsq %*% (x - mu)
+    h <- -isigsq
+    return (list(f = f, g = g, h = h))
+  }
\end{Sinput}
\end{Schunk}
We now draw 500 samples from this log-desity, using pre-specified values for \code{mu} (mean vector) and \code{isigsq} (inverse of the covariance matrix, or precision matrix) in a 3-dimensional state space:
\begin{Schunk}
\begin{Sinput}
R> K <- 3
R> mu <- runif(K, min = -0.5, max = +0.5)
R> isigsq <- matrix(runif(K*K, min = 0.1, max = 0.2), ncol = K)
R> isigsq <- 0.5*(isigsq + t(isigsq))
R> diag(isigsq) <- rep(0.5, K)
R> x.init <- rep(0.0, K)
R> x.smp <- sns.run(x.init, logdensity.mvg, niter = 500
+    , mh.diag = TRUE, mu = mu, isigsq = isigsq)
\end{Sinput}
\end{Schunk}
Next, we use the \code{summary.sns} function to view some of the diagnostics:
\begin{Schunk}
\begin{Sinput}
R> summary(x.smp)
\end{Sinput}
\begin{Soutput}
Stochastic Newton Sampler (SNS)
state space dimensionality:  3 
total iterations:  500 
	NR iterations:  10 
	burn-in iterations:  250 
	end iteration:  500 
	thinning interval:  1 
	sampling iterations (before thinning):  250 
acceptance rate:  1 
	mean relative deviation from quadratic approx: -1.64e-14 % (post-burnin)
sample statistics:
	(nominal sample size: 250)
         mean          sd         ess        2.5%         50%  97.5%
1   0.1003547   1.4953277 250.0000000  -2.7359187   0.0512754 3.0394
2   0.0752989   1.5091195 226.3462010  -2.7690970   0.1056509 3.0710
3   0.0518412   1.4314792 250.0000000  -2.7654234  -0.0088294 2.7638
  p-val
1 0.984
2 0.944
3 0.992
summary of ess:
   Min. 1st Qu.  Median    Mean 3rd Qu.    Max. 
  226.3   238.2   250.0   242.1   250.0   250.0 
\end{Soutput}
\end{Schunk}
As expected, the acceptance rate is 100\%, and there is no deviation from quadratic approximation, for SNS sampling of a multivariate Gaussian.

In real-world applications, the Gaussian proposal function is only an approximation for the sampled distribution (since log-density is not quadratic), creating the Markovian dependency and less-than-perfect acceptance rate. We study one such example next.

\subsection{Example 2: Bayesian Poisson regression}\label{subsection-example-poisson}
Generalized Linear Models (GLMs)~\citep{nelder1972generalized} are an important family of statistical models with applications such as risk analysis~\citep{sobehart2000moody}, public health~\citep{azar2011immunologic} and political science~\citep{gelman2007data}. GLMs can be extended to incorporate data sparseness and heterogeneity via the Hierarchical Bayesian framework~\citep{peter2005bayesian} or to account for repeated measurements and longitudinal data via Generalized Linear Mixed Model~\citep{mcculloch2006generalized}. With properly-chosen link functions, many GLMs are known - or can be easily proven - to have globally-concave log-densities with negative-definite Hessian matrices~\citep{gilks1992adaptive,mahani2015expander}. As such, GLMs are excellent candidates for SNS. Embedded in Bayesian frameworks, they continue to enjoy log-concavity assuming the same property holds for prior terms, according to the Bayes' rule and the invariance of concavity under function addition.

\begin{figure}
\includegraphics{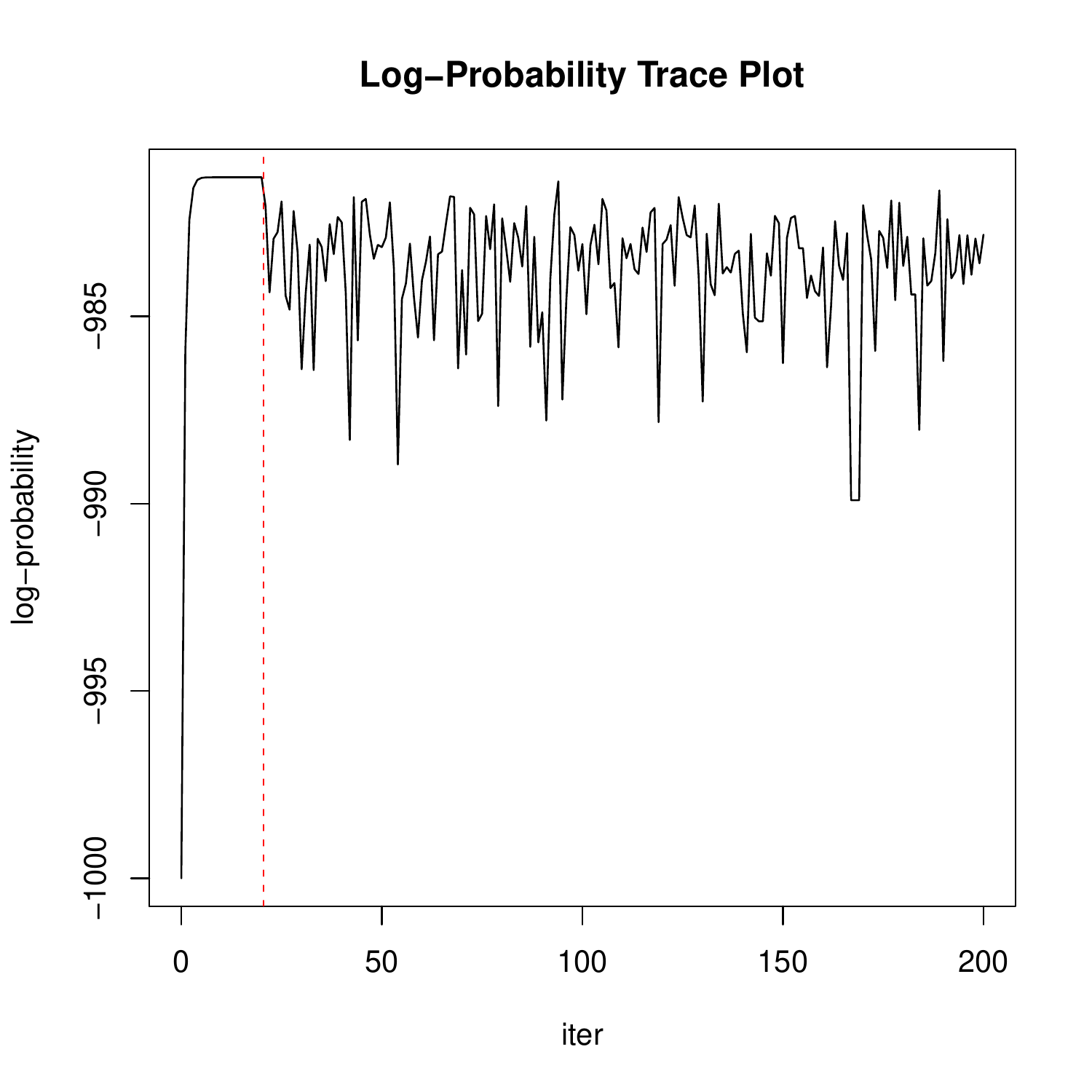}
\caption{Log-probability trace plot for Poisson regression problem with $K=5$ and $N=1000$. Vertical line separates non-stochastic mode (left, 20 iterations) from stochastic mode (right, 180 iterations).}
\label{fig-poisson-lp}
\end{figure}

In our second example, we illustrate how to apply \pkg{sns} to the log-likelihood of Poisson regression. As before, we start with constructing the log-density and its gradient and Hessian, using the expander framework of \pkg{RegressionFactory} package:
\begin{Schunk}
\begin{Sinput}
R> library(RegressionFactory)
R> loglike.poisson <- function(beta, X, y) {
+    regfac.expand.1par(beta, X = X, y = y
+      , fbase1 = fbase1.poisson.log)
+  }
\end{Sinput}
\end{Schunk}
Now we simulate data from the generative model:
\begin{Schunk}
\begin{Sinput}
R> K <- 5
R> N <- 1000
R> X <- matrix(runif(N * K, -0.5, +0.5), ncol = K)
R> beta <- runif(K, -0.5, +0.5)
R> y <- rpois(N, exp(X %*% beta))
\end{Sinput}
\end{Schunk}
For reference, we do a maximum-likelihood (ML) estimation of the coefficients using \code{glm} command:
\begin{Schunk}
\begin{Sinput}
R> beta.init <- rep(0.0, K)
R> beta.glm <- glm(y ~ X - 1, family = "poisson", start = beta.init)$coefficients
\end{Sinput}
\end{Schunk}
As mentioned before, \pkg{sns} can be run in non-stochastic mode, which is equivalent to Newton-Raphson optimization with line search. Results should be identical, or very close, to \code{glm} results:
\begin{Schunk}
\begin{Sinput}
R> beta.sns <- sns.run(beta.init, fghEval = loglike.poisson
+    , niter = 20, nnr = 20, X = X, y = y)
R> beta.nr <- beta.sns[20, ]
R> cbind(beta.glm, beta.nr)
\end{Sinput}
\begin{Soutput}
       beta.glm      beta.nr
X1  0.227484461  0.227484461
X2 -0.003779822 -0.003779822
X3  0.422692916  0.422692916
X4 -0.324037953 -0.324037953
X5  0.321018415  0.321018415
\end{Soutput}
\end{Schunk}
The primary use-case for \pkg{sns} is not ML estimation, but rather MCMC sampling of the distribution. To do this, we perform the first few iterations in non-stochastic mode (20 iterations here), and then switch to stochastic mode for the remaining 180 iterations:
\begin{Schunk}
\begin{Sinput}
R> beta.smp <- sns.run(beta.init, loglike.poisson
+    , niter = 200, nnr = 20, mh.diag = TRUE, X = X, y = y)
\end{Sinput}
\end{Schunk}
Examining the log-probability trace plot (Figure~\ref{fig-poisson-lp}) shows an expected pattern: During non-stochastic phase (left of vertical line) log-probability rises steadily while approaching the peak. During MCMC sampling, on the other hand, PDF maximum forms an upper bound for the MCMC movements, and the chain occasionally visits low-probability areas. The plot is created using the following line:
\begin{Schunk}
\begin{Sinput}
R> plot(beta.smp, select = 1)
\end{Sinput}
\end{Schunk}
The \code{plot.sns} function offers 4 other types of plots, besides the log-probability trace plot. We refer the reader to the package documentation for details. Further diagnostic information can be accessed via the \code{summary.sns} function:
\begin{Schunk}
\begin{Sinput}
R> summary(beta.smp)
\end{Sinput}
\begin{Soutput}
Stochastic Newton Sampler (SNS)
state space dimensionality:  5 
total iterations:  200 
	NR iterations:  20 
	burn-in iterations:  100 
	end iteration:  200 
	thinning interval:  1 
	sampling iterations (before thinning):  100 
acceptance rate:  0.95 
	mean relative deviation from quadratic approx: 0.421 % (post-burnin)
sample statistics:
	(nominal sample size: 100)
         mean          sd         ess        2.5%         50%   97.5%
1   0.2237480   0.1138531  59.6866795   0.0170658   0.2145079  0.4461
2   0.0071665   0.0941524 100.0000000  -0.1866930   0.0262237  0.1786
3   0.4157330   0.1155381 100.0000000   0.1765806   0.4109959  0.6370
4  -0.3115131   0.1226679 100.0000000  -0.4931933  -0.3332426 -0.0110
5   0.3110794   0.1010334 100.0000000   0.1190488   0.3240565  0.4663
  p-val   
1  0.04 * 
2  0.90   
3  0.01 **
4  0.06 . 
5  0.01 **
---
Signif. codes:  0 '***' 0.001 '**' 0.01 '*' 0.05 '.' 0.1 ' ' 1
summary of ess:
   Min. 1st Qu.  Median    Mean 3rd Qu.    Max. 
  59.69  100.00  100.00   91.94  100.00  100.00 
\end{Soutput}
\end{Schunk}
The \code{summary.sns} function discards the first half of the samples as burn-in by default, before calculating sample statistics and acceptance. This behavior can be controlled via the argument \code{nburnin}. Arguments \code{end} and \code{thin} have behavior similar to their counterparts in the \code{as.mcmc} function of \pkg{coda} package. We observe two numbers in the summary print-out: Firstly, acceptance rate is less than 100\%, contrary to the case with a multivariate Gaussian PDF (example 1). Secondly, the mean relative deviation from quadratic approximation (\code{reldev.mean}) is now non-zero, again reflecting non-Gaussianity of the poisson likelihood PDF. The number (<1\%) is still small, however, leading to high acceptance rate and good mixing of the chain.

Next, we want to predict the response variable in Poisson regression model, given new values of the explanatory variables. We distinguish between two types of prediction: 1) predicting mean response, 2) generating samples from posterior predictive distribution. We illustrate how to do both using the \code{predict.sns} function. We begin by implementing the mean prediction function, which simply applies the inverse link function (exponential here) to the linear predictor. (For better comparison between the two prediction modes, we increase number of samples to 1000)
\begin{Schunk}
\begin{Sinput}
R> beta.smp <- sns.run(beta.init, loglike.poisson
+    , niter = 1000, nnr = 20, mh.diag = TRUE, X = X, y = y)
R> predmean.poisson <- function(beta, Xnew) exp(Xnew %*% beta)
\end{Sinput}
\end{Schunk}
The following single line performs sample-based prediction of mean response (using \code{X} in lieu of \code{Xnew} for code brevity):
\begin{Schunk}
\begin{Sinput}
R> ymean.new <- predict(beta.smp, predmean.poisson
+    , nburnin = 100, Xnew = X)
\end{Sinput}
\end{Schunk}
\code{ynew} is a matrix of \code{N} (1000) rows and \code{niter - nburnin} (900) columns. Each row corresponds to an observation (one row of \code{Xnew}), and each column corresponds to a prediction sample (one row of \code{beta.smp} after burn-in).

We can also generate samples from posterior predictive distribution as follows:
\begin{Schunk}
\begin{Sinput}
R> predsmp.poisson <- function(beta, Xnew)
+    rpois(nrow(Xnew), exp(Xnew %*% beta))
R> ysmp.new <- predict(beta.smp, predsmp.poisson
+    , nburnin = 100, Xnew = X)
\end{Sinput}
\end{Schunk}
Comparing prediction summaries is illuminating:
\begin{Schunk}
\begin{Sinput}
R> summary(ymean.new)
R> summary(ysmp.new)
\end{Sinput}
\end{Schunk}
\begin{Schunk}
\begin{Soutput}
prediction sample statistics:
	(nominal sample size: 900)
        mean         sd        ess       2.5%        50%  97.5%
1   0.865031   0.064676 900.000000   0.742812   0.862279 1.0019
2   1.267430   0.106293 900.000000   1.077680   1.259413 1.4882
3   1.405566   0.073465 731.714663   1.261902   1.403256 1.5501
4   0.889331   0.063354 900.000000   0.773876   0.887845 1.0236
5   1.154694   0.081166 900.000000   1.009981   1.150145 1.3199
6   1.206132   0.076625 808.257228   1.071271   1.202108 1.3671
...
\end{Soutput}
\begin{Soutput}
prediction sample statistics:
	(nominal sample size: 900)
       mean        sd       ess      2.5%       50% 97.5%
1   0.87333   0.93240 900.00000   0.00000   1.00000     3
2   1.28000   1.15218 900.00000   0.00000   1.00000     4
3   1.37444   1.19640 900.00000   0.00000   1.00000     4
4   0.87556   0.93864 900.00000   0.00000   1.00000     3
5   1.15000   1.02296 526.62253   0.00000   1.00000     3
6   1.23000   1.10987 900.00000   0.00000   1.00000     4
...
\end{Soutput}
\end{Schunk}

In the limit of infinite samples, the mean predictions from the two methods will be equal, and they are quite close based on 900 samples above. However, standard deviation of predictions is much larger for \code{predsmp.poisson} compared to \code{predmean.poisson}, as the former combines the uncertainty of coefficient values (represented in the \code{sd} values for \code{beta}'s) with the uncertainty of samples from Poisson distribution around the mean, i.e. the \code{sd} of Poisson distribution. Also note that, as expected, quantiles for \code{predsmp.poisson} are discrete since predictions are discrete, while the quantiles for \code{predmean.poisson} are continuous as predictions are continuous in this case.

\subsection{Example 3: High-dimensional Bayesian Poisson regression}\label{subsection-example-highD}
Contrary to standard Metropolis variants with multivariate Gaussians centered on current point, SNS is an aggressive, non-local MCMC algorithm as it seeks to construct a global, Gaussian approximation of the PDF. Under favorable conditions, this can lead to uncorrelated chains and efficient sampling, with the extreme case of perfectly uncorrelated samples for a multivariate Gaussian distribution. An important pathological case arises when the state space dimensionality is high. Continuing with the Poisson regression example, we increase $K$ from 5 to 100, while holding $N=1000$. To illustrate that the problem is not covergence but mixing, we explicitly use the \code{glm} estimate (mode of PDF) as the initial value for the MCMC chain:
\begin{Schunk}
\begin{Sinput}
R> K <- 100
R> X <- matrix(runif(N * K, -0.5, +0.5), ncol = K)
R> beta <- runif(K, -0.5, +0.5)
R> y <- rpois(N, exp(X %*% beta))
R> beta.init <- glm(y ~ X - 1, family = "poisson")$coefficients
R> beta.smp <- sns.run(beta.init, loglike.poisson
+    , niter = 100, nnr = 10, mh.diag = TRUE, X = X, y = y)
R> summary(beta.smp)
\end{Sinput}
\end{Schunk}
\begin{Schunk}
\begin{Soutput}
Stochastic Newton Sampler (SNS)
state space dimensionality:  100 
total iterations:  100 
	NR iterations:  10 
	burn-in iterations:  50 
	end iteration:  100 
	thinning interval:  1 
	sampling iterations (before thinning):  50 
acceptance rate:  0.16 
	mean relative deviation from quadratic approx: 8.35 % (post-burnin)
sample statistics:
	(nominal sample size: 50)
       mean        sd       ess      2.5%       50%   97.5% p-val  
1  0.421594  0.109166  9.760501  0.164051  0.447386  0.5166  0.02 *
2  0.219483  0.105112  4.913961 -0.081762  0.223558  0.4021  0.12  
3  0.203045  0.078498  5.749257  0.060661  0.250520  0.3404  0.02 *
4 -0.100247  0.069272  2.173084 -0.249282 -0.066971 -0.0368  0.04 *
5  0.111424  0.064391  4.121670  0.032764  0.122196  0.2168  0.02 *
6  0.115763  0.106802  5.238306 -0.064935  0.089167  0.2260  0.24  
---
Signif. codes:  0 '***' 0.001 '**' 0.01 '*' 0.05 '.' 0.1 ' ' 1
...
summary of ess:
   Min. 1st Qu.  Median    Mean 3rd Qu.    Max. 
  2.096   4.095   5.335   6.991   6.694  72.500 
\end{Soutput}
\end{Schunk}

We see a significant drop in acceptance rate as well as effective sample sizes for the coefficients. Also note that mean relative deviation from quadratic approximation is now nearly 10x larger than the value for $K=5$. To improve mixing, we use the `state space partitioning' strategy of \pkg{sns}, available via the \code{part} argument of \code{sns} and \code{sns.run}. This leads to SNS sampling of subsets of state space wrapped in Gibbs cycles, with each subset being potentially much lower-dimensional than the original, full space. This strategy can significantly improve mixing. Below we use the function \code{sns.make.part} to partition the 100-dimensional state space into 10 subsets, each 10-dimensional:
\begin{Schunk}
\begin{Sinput}
R> beta.smp.part <- sns.run(beta.init, loglike.poisson
+    , niter = 100, nnr = 10, mh.diag = TRUE
+    , part = sns.make.part(K, 10), X = X, y = y)
R> summary(beta.smp.part)
\end{Sinput}
\end{Schunk}
\begin{Schunk}
\begin{Soutput}
Stochastic Newton Sampler (SNS)
state space dimensionality:  100 
state space partitioning:  10  subsets
total iterations:  100 
	NR iterations:  10 
	burn-in iterations:  50 
	end iteration:  100 
	thinning interval:  1 
	sampling iterations (before thinning):  50 
acceptance rate:  0.94 
sample statistics:
	(nominal sample size: 50)
        mean         sd        ess       2.5%        50%  97.5% p-val
1  0.3969166  0.0885804 39.1691610  0.2100833  0.3964180 0.5402  0.02
2  0.1896767  0.1152469 50.0000000 -0.0248610  0.1865183 0.3670  0.08
3  0.1579658  0.0947398 28.9506128 -0.0122672  0.1549021 0.3189  0.08
4 -0.1147680  0.0980224 29.6244795 -0.2697117 -0.1198215 0.0670  0.32
5  0.1540473  0.1003371 50.0000000 -0.0239445  0.1473975 0.3207  0.12
6  0.1466464  0.1027434 32.5836802 -0.0095312  0.1411849 0.3683  0.08
   
1 *
2 .
3 .
4  
5  
6 .
---
Signif. codes:  0 '***' 0.001 '**' 0.01 '*' 0.05 '.' 0.1 ' ' 1
...
summary of ess:
   Min. 1st Qu.  Median    Mean 3rd Qu.    Max. 
  12.77   28.89   50.00   41.59   50.00   93.70 
\end{Soutput}
\end{Schunk}

\begin{figure}[H]
\vspace{6pc}
\includegraphics[scale=0.75]{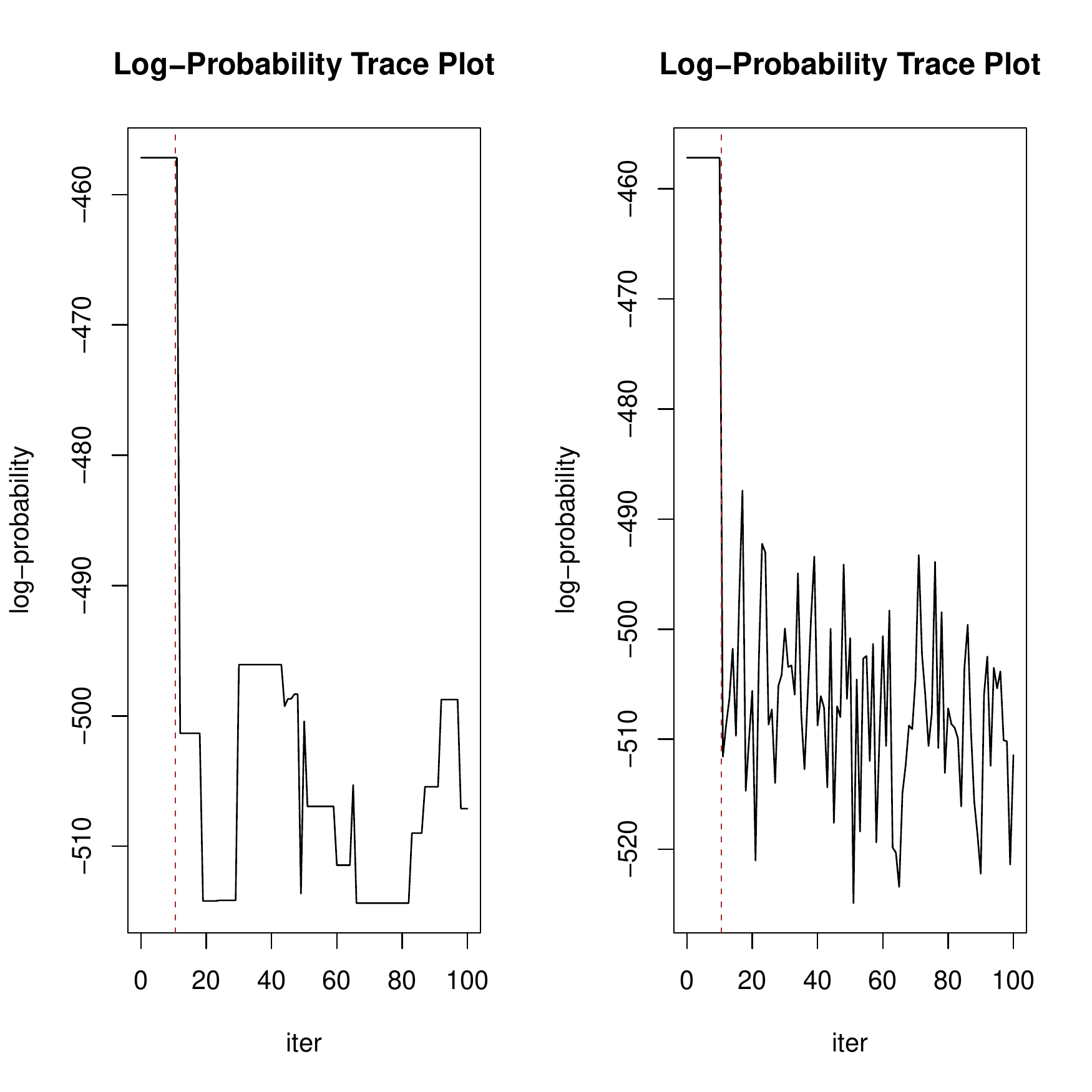}
\caption[]{Comparison of log-probability trace plots for $N=1000$ and $K=100$, without (left) and with (right) state space partitioning using 10 subsets.}
\label{fig-poisson-lp-sbs}
\end{figure}

Notice the improved acceptance rate as well as effective sample sizes. A comparison of log-probability trace plots confirms better mixing after convergence to PDF mode (see Figure~\ref{fig-poisson-lp-sbs}).
\begin{Schunk}
\begin{Sinput}
R> par(mfrow = c(1,2))
R> plot(beta.smp, select = 1)
R> plot(beta.smp.part, select = 1)
\end{Sinput}
\end{Schunk}


\section{Summary}\label{section-summary}
In this paper we presented \pkg{sns}, an \proglang{R} package for Stochastic Newton Sampling of twice-differentiable, log-concave PDFs, where a multivariate Gaussian resulting from second-order Taylor series expansion of log-density is used as proposal function in a Metropolis-Hastings framework. Using an initial non-stochastic mode, equivalent to Newton-Raphson optimization with line search, allows the chain to rapidly converge to high-density areas, while `state space partitioning', Gibbs sampling of full state space in lower-dimensional blocks, allows SNS to overcome mixing problems while sampling from high-dimensional PDFs. There are several opportunities for further research and development.

\textbf{Beyond twice-differentiability and log-concavity:} Current version of SNS requires the log-density to be twice-differentiable and concave. In many real-world application, the posterior PDF does not have a negative-definite Hessian, or it cannot be proven to have such a property. In such cases, SNS would not be applicable to the entire PDF. However, if one can identify blocks within the full Hessian that does enjoy such property, then SNS can be combined with other, more generic sampling algorithms such as slice sampler or HMC, all embedded in a Gibbs cycle. The implementation would be similar to that of state space partitioning approach in \pkg{sns}, but replacing SNS with alternative samplers for some subsets. An alternative approach is discussed in~\cite{geweke2001bayesian} in the context of state-space models, where non-concave cases are handled by utilizing exponential and uniform distributions. Convergence and mixing properties of such extensions to general cases must be carefully studied.

An important situation where twice-differentiability is violated is in the presence of boundary conditions. The current SNS algorithm assumes unconstrained state space. One way to deal with constrained subspaces is, again, to mix and match SNS and other samplers within Gibbs framework. For example, the slice sampler algorithm implemented in \pkg{MfUSampler}~\citep{mahani2014mfusampler} is capable of dealing with boxed constraints. It can therefore be assigned to subspaces with constraints, and the unconstrained subspaces can be handled by SNS. Further research is needed in order to relax twice-differentiability and log-concavity requirements for SNS.

\textbf{Optimized state space partitioning:} While our experiments suggest that the optimal number of dimensions within each subset is around 10, yet more comprehensive analysis, both theoretical and empirical, is needed to determine when/how to partition the state space. The correlation structure of the state space might need to be taken into account while assigning coordinates to subsets, as well as data dimensions such as number of observations. Our preliminary, uni-dimensional analysis shows that mixing improves with $N^{1/2}$, where $N$ is the number of independently-sampled observations in a regression problem.

Computational implications of subset size require more attention as well. In particular, the current implementation in \pkg{sns} calls the same underlying, full-space \code{fghEval} function. This means within each subset, full Hessian is calculated, which is computationally wasteful. This naive approach neutralizes a potential computational advantage of state space partitioning, i.e., reducing the cost of Hessian evaluation which goes up quadratically with state space dimensionality. The tradeoff is a more complex API for function evaluation, which is the responsibility of the user.

\textbf{Performance benchmarking:} The ultimate test of an MCMC algorithm is the speed with which it can generate independent samples from a PDF. SNS often generates MCMC chains that are significantly less correlated than univariate techniques, or even Metropolis with Gaussian proposal (centered on current point). On the other hand, the computational cost of generating a sample for SNS is often higher due to the expensive step of calculating the Hessian. Our experiments show that the next effect is such that SNS becomes the most efficient technique when applicable (i.e., when Hessian is negative-definite). However, a fair comparison of sampling techniques - in a way that is as independent from a particular hardware, software and implementation as possible - is non-trivial. \cite{thompson2011compare} suggests the 'number of function evaluations per independent sample' as a metric for comparing samplers. This metric must be extended to allow for comparison of samplers that utilize the gradient vector and/or the Hessian matrix, such as SNS. Expressing the computational cost of gradient and Hessian evaluation in the `currency' or units of function evaluation seems as the logical extension of the metric.

\bibliography{SNS.bib}
\end{document}